\documentclass[useAMS,usenatbib,referee]{biom}
\def\bSig\mathbf{\Sigma}

\usepackage{amsmath}

\usepackage[figuresright]{rotating}
\usepackage{url}

\title{Potential outcome simulation for efficient head-to-head comparison of adaptive dose-finding designs}

\author
{Michael Sweeting$^{1,*}$\emailx{michael.sweeting@astrazeneca.com},
Daniel Slade$^{2}$, Dan Jackson$^{1}$, 
and
Kristian Brock$^{3}$ \\
$^{1}$ Statistical Innovation, Oncology Biometrics, AstraZeneca, UK \\
$^{2}$ Early Oncology Statistics, Oncology Biometrics, AstraZeneca, UK \\
$^{3}$ Statistical Innovation, Data Sciences and AI, AstraZeneca, UK
}

\begin{document}

\label{firstpage}

\begin{abstract}
Dose-finding trials are a key component of the drug development process and rely on a statistical design to help inform dosing decisions. Triallists wishing to choose a design require knowledge of operating characteristics of competing methods. This is often assessed using a large-scale simulation study with multiple designs and configurations investigated, which can be time-consuming and therefore limits the scope of the simulation.

We introduce a new approach to the design of simulation studies of dose-finding trials. The approach simulates all potential outcomes that individuals could experience at each dose level in the trial. Datasets are simulated in advance and then the same datasets are applied to each of the competing methods to enable a more efficient head-to-head comparison. 

In two case-studies we show sizeable reductions in Monte Carlo error for comparing a performance metric between two competing designs. Efficiency gains depend on the similarity of the designs. Comparing two Phase I/II design variants, with high correlation of recommending the same optimal biologic dose, we show that the new approach requires a simulation study that is approximately 30 times smaller than the conventional approach. Furthermore, advance-simulated trial datasets can be reused to assess the performance of designs across multiple configurations.

We recommend researchers consider this more efficient simulation approach in their dose-finding studies and we have updated the R package \texttt{escalation} to help facilitate implementation.

\end{abstract}


\maketitle

\section{Introduction}
Phase I clinical trials aim to assess the safety of a novel compound and to identify a maximum tolerated dose (MTD) through dose escalation \citep{storer_design_1989, oquigley_continual_1990}. Phase I/II clinical trials meanwhile aim to use information on both toxicity and an efficacy response to find through dose-escalation an optimal dosage of a drug which has high effectiveness as well as tolerable toxicity \citep{yin_bayesian_2006}. Frequently such trials use adaptive dose finding designs to reach their objective, where patients are recruited in cohorts and the dosage for the next cohort is escalated, deescalated or unchanged according to the inferences made on doses and outcomes from the previous cohorts.

A wide range of dose-finding methods exist for Phase I and Phase I/II trial design and triallists are often faced with a difficult decision of which to use for their early phase study. To understand the operating characteristics of different designs, large-scale simulation studies are usually performed. However, there are often marginal performance gains seen in the comparison of competing designs, and it is not always clear whether differences are true performance differences or due to chance variation. Most dose-finding methods require fine-tuning of key components of the design, such as the sample and cohort size, safety modifications and stopping rules - all of which can lead to small differences in model performance. Comprehensively assessing such components is a recommended part of the design process to assist the trial team to select a final design \citep{wheeler_how_2019}. However, this can be a time-consuming process as each simulation must be large enough to reduce Monte Carlo (MC) simulation error.

Outside of dose-finding designs, simulation studies often contrast competing statistical methods by applying each method to the same simulated datasets. This recommended approach reduces MC error for between-method contrasts. The MC error will affect the methods' performance in the same way since methods are matched on the same generated data \citep{morris_using_2019}. For example, by using the same simulated datasets for two competing methods we are asking, for each simulated dataset, the counterfactual question of what the difference is in performance had we used the second method instead of the first method on the \emph{same study population}.

In an adaptive trial, such as a dose-finding study, generating the same simulated data for competing methods is more challenging because the method itself determines the data that is to be collected. Two methods for dose-finding may deviate in their recommendation of the dose that the next cohort of patients who enter the trial should receive, and this has consequences for the outcome that is subsequently simulated. Because of this, simulations of dose-finding studies have typically simulated data as the trial progresses, dependent on the decisions being made.

This manuscript approaches simulation in dose-finding studies in a novel way. We propose to simulate data ahead of applying the competing methods and we do this by considering all the \emph{potential outcomes} (POs) for each individual that could be recruited into the trial. Potential outcomes are the outcomes at each dose level that would be observed if the design chose to assign the individual at that given dose level. PO simulations have been used by \cite{oquigley_non-parametric_2002} and \cite{cheung_simple_2014} for calculating a non-parametric optimal benchmark to use as a gold standard for the upper performance limit to compare with the performance of a method. In this manuscript we are recommending using PO simulated data for a different purpose, namely to apply competing dose-finding designs to the same datasets, where each design only observes a partial outcome profile. We show that this simple rethinking of simulation studies in adaptive trials can lead to substantial gains in reducing MC error for between-method contrasts, and hence can increase computational efficiency and lead to smaller simulation studies. 

To facilitate implementation of the new PO simulation approach, we have programmed the method as the default option within the \texttt{escalation} \citep{escalation} package in \texttt{R}. More generally, we recommend these approaches are applied widely and incorporated into other existing software for Phase I trial designs. Ideally, software should be flexible enough to allow existing PO simulated datasets to be passed to it to assess performance of the design on pre-existing simulated data. 

The structure of this manuscript is as follows.
In Section \ref{s:po_in_phase1}, we introduce potential outcomes from the point of view of a dose-finding clinical trial that investigates binary toxicity and/or efficacy outcomes.
In Section \ref{s:sim_po}, we introduce methods for simulating potential outcomes under different response assumptions. 
In Section \ref{s:case_study1}, we illustrate our approach in a phase I/II setting with co-primary toxicity and efficacy outcomes, to hone the dose-exclusion parameters of a BOIN12 \citep{lin_boin12_2020} design.
In Section \ref{s:case_study2}, we illustrate our approach in a comparison of two common Phase I dose findings designs; namely the continual reassessment method (CRM) and the modified toxicity probability interval (mTPI-2) design.
In Section \ref{s:implementation} we discuss implementation of the PO approach in the \texttt{escalation} R package.
Finally in Section \ref{s:discussion}, we conclude with some discussion.

\section{Potential Outcomes in Phase I trials}
\label{s:po_in_phase1}
A potential outcome is the outcome that would occur if the patient receives a certain dose, where the potential outcome terminology comes from the Neyman-Rubin causal inference framework \citep{rubin_causal_2005}. In Phase I dose-finding studies, a binary toxicity outcome (dose-limiting toxicity (DLT) or no DLT) is usually measured. To simulate potential toxicity outcomes at the individual level we can take advantage of the often-made assumption that a DLT outcome is monotonically non-decreasing in dose. In words, this says that if a patient had a DLT at dose level $d$ then they would also have had a DLT had they been given dose level $d+1$, or dose level $d+2$, etc. This is a defensible assumption that has been made in most Phase I dose-finding studies to date \citep{oquigley_experimental_2006} and gives rise to monotonically non-decreasing toxicity probabilities at the aggregate level. Specifically, O’Quigley and Zohar state that monotonicity “… means that when a patient experienced a dose-limiting toxicity at a specific level, then, had this same patient been treated at any higher level he would also have suffered dose-limiting toxicity. Conversely, were the patient to tolerate the treatment at a specific dosage, then, for all lower levels, the patient would also have tolerated treatment.” \citep{oquigley_experimental_2006} A simulated PO dataset with monotonically non-decreasing DLT outcomes will ensure that if Method A doses patient $i$ at dose level 2, say, and a DLT is observed, then Method B that chooses a dose level greater than or equal to 2 for patient $i$ will also observe a DLT for this patient. 

Table \ref{tab:po_example} demonstrates PO toxicity data simulated for ten patients. The true probabilities of DLT from which the POs are simulated are given at the top of the table. At the individual level, the POs are monotonically non-decreasing and the probability of DLT calculated from aggregating the individual level POs is also monotically non-decreasing. The data is then used as a simulated dataset for a dose-esclation study. In this example we assume patients enter the trial in cohort sizes of 2 (so that decisions on whether to escalate or not are made after every 2 patients). This fictional design doses the first cohort at dose level 1, and the outcomes observed are a DLT ($Y=1$) and no-DLT ($Y=0$) for patients 1 and 2, respectively. The observed outcomes are marked by a * in Table \ref{tab:po_example}. Cohort 2 stays at dose level 1, whilst cohort 3 is given dose level 2, cohort 4 is given dose level 3 and cohort 5 is given dose level 4. The observed data is therefore the realisation of the POs at the given dose levels and is summarised at the bottom of Table \ref{tab:po_example}. Note that the empirical probabilities of DLT from the observed data need not be monotonically non-decreasing, as seen in this example.

\begin{table}
\caption{\label{tab:po_example} Example of potential toxicity outcome data simulated for 10 patients, and the outcomes observed (marked by a *) when conducting a dose-finding study.}
\begin{center}
\begin{tabular}{cccccc}
& \multicolumn{5}{c}{Dose Level} \\
& 1 & 2 & 3 & 4 & 5 \\
True p(DLT) & 0.10 & 0.20 & 0.22 & 0.35 & 0.50 \\
\hline 
\multicolumn{6}{c}{Potential Outcomes} \\
Patient 1 & 1* & 1 & 1 & 1 & 1\\
Patient 2 & 0* & 1 & 1 & 1 & 1\\
\hline 
Patient 3 & 0* & 0 & 0 & 0 & 1\\
Patient 4 & 0* & 0 & 0 & 0 & 0\\
\hline 
Patient 5 & 0 & 0* & 0 & 0 & 0\\
Patient 6 & 0 & 0* & 0 & 1 & 1\\
\hline 
Patient 7 & 0 & 0 & 0* & 0 & 0\\
Patient 8 & 0 & 0 & 0* & 0 & 0\\
\hline 
Patient 9 & 0 & 0 & 0 & 0* & 0\\
Patient 10 & 0 & 0 & 1 & 1* & 1\\
\hline 
PO empirical p(DLT) & 0.10 & 0.20 & 0.30 & 0.40 & 0.60 \\
\hline \\
Data ($y/n$) & 1/4 & 0/2 & 0/2 & 1/2 & - \\
Empirical p(DLT) & 0.25 & 0.00 & 0.00 & 0.50 & -
\end{tabular}
\end{center}
\end{table}

If the outcome is an efficacy measure (e.g. response or no response) then we may or may not be willing to make the same assumption that outcome is monotonically non-decreasing in dose at the individual level. By making this assumption we are saying that if a patient had a response at dose level $d$ then they would also have had a response had they been given dose level $d+1$, or dose level $d+2$, etc…   The true objective response rates (ORR) at the aggregate level are then monotonically non-decreasing. To note, this assumption also accommodates the scenario where ORRs plateau and stay that same after a given dose. However, at the individual level, it may be more plausible that a response at dose $d$ is not guaranteed to give a response at dose level $d+1$, perhaps because of tolerability issues at higher doses. In this case, potential efficacy outcomes can still be simulated by relaxing the strict monotonically non-decreasing assumption.

\section{Simulating potential outcomes}
\label{s:sim_po}
Within a Phase I or I/II dose-finding study, assume there are a maximum of $N$ patients to be recruited. Suppose the objective for the statistician is to assess the operating characteristics of key quantities (e.g. amount of underdosing, target dosing, overdosing, final recommended dose, maximum sample size), of two possible competing designs that provide dosing decisions to a set of discrete dose levels using dose-escalation. The comparison of the performance of two designs on the same patients is possible in a simulation study using the potential outcome framework.

\subsection{Simulating potential toxicity outcomes under monotonicity}\label{section:sim_monotonic_outcomes}
To simulate toxicity binary outcomes (DLTs) that are monotonically non-decreasing with dose consider the set of true DLT probabilities for the $D$ doses under consideration $\pi_T(d), d=1,\ldots,D$, where subscript $T$ refers to the toxicity endpoint. These DLT probabilities are monotonically non-decreasing such that $\pi_T(1) \le \pi_T(2) \le \pi_T(3) \le \ldots \le \pi_T(D)$. 
Traditionally, in a dose-finding simulation study, when a patient is recruited we might simulate an outcome for them depending on the dose allocated to them by the design. For example, if they were allocated dose $d$ then we would simulate for patient $i$, 	$Y_{T,i} \sim \textrm{Binomial}(1,\pi_T(d))$. The issue with this simulation approach is that we can only simulate this outcome during the conduct of the trial since we don’t know from the outset which dose the patient will receive. The design makes the dosing choice for patient $i$ based on all accumulated evidence up to patient $i$. This simulation approach has implications when comparing to a second design. For example, the first design might recommend patient $i$ to receive dose $k$ and we simulate a DLT event $Y_{T,i}=1$ for this patient based on a draw from the $\textrm{Binomial}(1,\pi_T(k))$ distribution. The second design might instead recommend patient $i$ receives dose $k+1$, and we might simulate no DLT event ($Y_{T,i}=0$) under this scenario based on a draw from the $\textrm{Binomial}(1,\pi_T(k+1))$ distribution. There is no guarantee that the two simulated outcomes for the two different designs are coherent at the individual level, and this can introduce unwanted stochastic variation (MC error) when comparing operating characteristics between designs.

Over many thousands of simulated trials, this unwanted stochastic variation will become less important; however in general it will take more simulations to ensure the Monte Carlo error for the contrast of interest is small enough to ignore. 
A more efficient simulation strategy is to simulate POs for each individual ahead of the hypothetical trial commencing. The approach to simulate toxicity POs under a monotonicity assumption proceeds as follows:

\begin{enumerate}
\item	Before the hypothetical trial begins, for each potential patient $i=1,\ldots,N$ that could be recruited into the trial (where $N$ is the maximum sample size), we generate $N$ uniform(0,1) random variables:
$u_{T,i} \sim U(0,1);  i=1,\ldots,N.$
\item Given the true toxicity probabilities $\pi_T(d), \; k=d,\ldots,D$, where $D$ is the maximum number of dose levels, the PO for patient $i$ at dose level $d$, labelled $Y_{T,i}(d)$, is equal to 1 if $\pi_T(d) \ge u_{T,i}$ and 0 otherwise. If the true toxicity probabilities are monotonically non-decreasing then the POs will likewise be monotonically non-decreasing.
\end{enumerate}	

As an example, suppose $u_{T,i}=0.3$ and the true DLT probabilities for 5 doses are $${0.05,0.10,0.25,0.40,0.60}.$$  Then for this patient the lowest dose at which they would observe a DLT outcome is dose level 4 since this is the lowest dose for which the true DLT probability is greater than their simulated $u_{T,i}$. Patient $i$’s POs are $\bm{Y}_{T,i} = \{0,0,0,1,1\}$, indicating that the patient would have a DLT outcome at dose levels 4 or 5 had they been dosed there, and no DLT outcomes at dose levels 1, 2, or 3. It can be shown that the expected value of the POs are equal to the true toxicity probabilities $\pi_T(d)=E[Y_T(d)]$ \citep{cheung_simple_2014}.

\subsection{Simulating potential efficacy outcomes under a general dose-response shape}\label{section:sim_umbrella_outcomes}
If we have a binary efficacy outcome such as response (yes or no), the individual POs may not necessarily be monotonically non-decreasing. Instead a more general dose-response shape may be more appropriate. The approach above can still be used to generate POs under a more general dose-response model based on the specification of the true probabilities of efficacy at each dose level.
\begin{enumerate}
\item 	Simulate for each patient $u_{E,i} \sim U(0,1);  i=1,\ldots,N$.
\item Given the true efficacy probabilities $\pi_E(d), \; k=d,\ldots,D$, the efficacy PO for patient $i$ at dose level $d$, labelled $Y_{E,i}(d)$, is equal to 1 if $\pi_E(d) \ge u_{E,i}$ and 0 otherwise.
\end{enumerate}

\begin{figure}
  \caption{\label{fig:example_umbrella_po_simulation} Example generation of potential efficacy outcomes where there is a single turning point in the efficacy probabilities with increasing dose.}
	\centering
  \includegraphics[width=\textwidth]{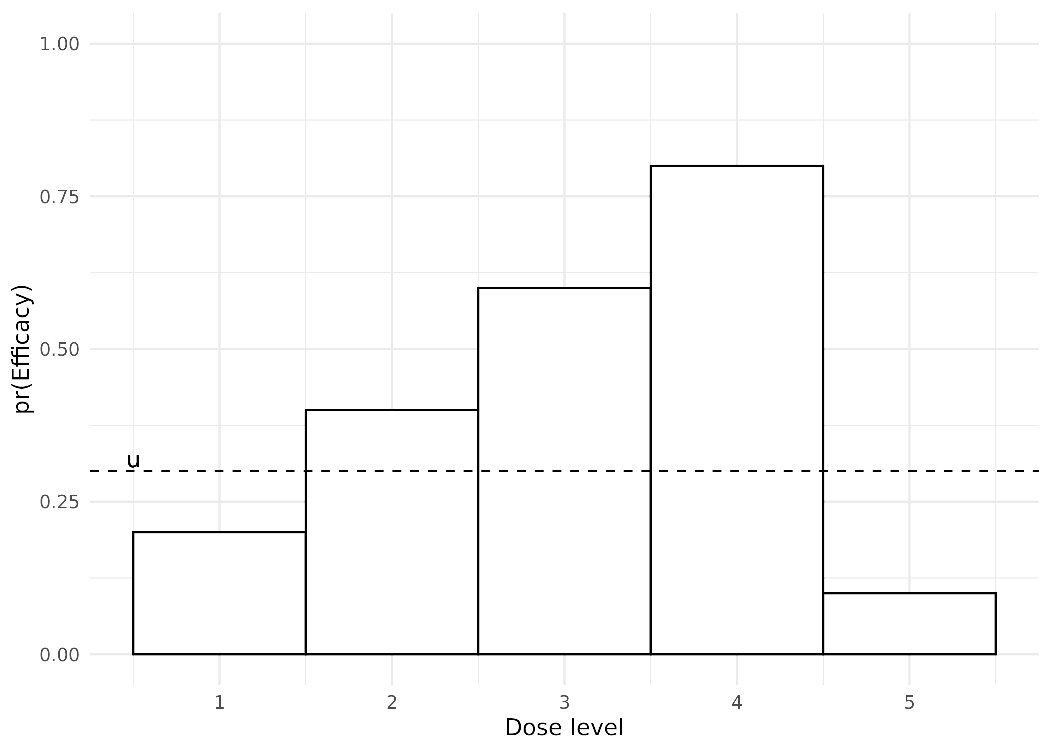}
\end{figure}

A graphical example of this algorithm is demonstrated in Figure \ref{fig:example_umbrella_po_simulation} where an umbrella shaped dose-response is envisaged. Suppose we simulate $u_{E,i}=0.3$ and the true efficacy probabilities for 5 doses are (0.2, 0.4, 0.6, 0.8, 0.1). Then the minimum effective dose is dose level 2 (the first dose above the horizontal dashed line in the plot). The minimum dose that no longer achieves efficacy is dose level 5 (the first dose after dose level 2 that is below the horizontal dashed line). We think of $u_{E,i}$ as an individual’s propensity for non-efficacy. A low $u_{E,i}$ means that all dose levels are likely to be effective for the individual whilst a high $u_{E,i}$ means that few or no dose levels are likely to be effective for the individual.

\subsection{Simulation of correlated potential toxicity and efficacy outcomes}
\label{section:sim_correlated_pos}
In Phase I/II designs, both toxicity and efficacy endpoints may be observed and used for escalation decisions. To generate correlated binary toxicity and efficacy PO data, we propose to invert latent correlated bivariate Gaussian variables. Our proposal proceeds as follows:

\begin{enumerate}
\item {\bf{Simulate correlated Gaussians}}. Simulate a pair of standard bivariate Gaussians with correlation $\rho$ for each patient:
$$
\begin{pmatrix}
\theta_{T,i} \\
\theta_{E,i}
\end{pmatrix}\sim N\left( \begin{pmatrix}
0 \\
0
\end{pmatrix},
\begin{pmatrix}
1 & \rho \\
\rho & 1
\end{pmatrix}
\right)
$$
\item {\bf{Derive correlated uniforms}}. Derive percentiles from the standard univariate Normal distribution, $u_{i,T} = \Phi(\theta_{i,T})$ and $u_{i,E} = \Phi(\theta_{i,E})$. $u_{i,T}$ and $u_{i,E}$ will be marginally uniformly distributed on (0, 1) with correlation dependent on $\rho$.
\item {\bf{Derive correlated binary outcomes}}. We use $u_{T,i}$ and $u_{E,i}$ to infer $Y_{T,i}(d)$ and $Y_{E,i}(d)$ following the PO approaches described in Sections (\ref{section:sim_monotonic_outcomes}) and (\ref{section:sim_umbrella_outcomes}).
\end{enumerate}

Note that using this approach, the Pearson correlation between the binary variables varies with the event probabilities and is not, in general, equal to $\rho$.

\section{Case study 1: Fine-tuning of a BOIN12 Phase I/II design}
\label{s:case_study1}

\subsection{BOIN12 Variants}
To illustrate the new simulation approach we apply the method to a case-study where two variants of the BOIN12 Phase I/II design \cite{lin_boin12_2020} are to be compared. BOIN12 is an extension of the BOIN dose-finding design \cite{yuan2016bayesian} that aims to find the optimal biologic dose (OBD) using both binary toxicity (DLT) and efficacy (response) data. The design uses a utility function to reflect the toxicity-efficacy trade-off and targets doses with the highest posterior mean utility. The design has many components that require fine-tuning. For further information on the intracacies of this design readers should refer to Yuan et al. \cite{yuan2016bayesian}. Two components of the design that require consideration are the dose-acceptability criteria used by BOIN12 to adaptively decide which doses may be used to treat the next cohort of patients. Doses are admissible only if there is not strong evidence that they are unsafe, such that $p(\pi_T(d) > \phi_T | \textrm{data}) < c_T$, where $\phi_T$ is an upper limit for toxicity and $c_T$ is a safety threshold, usually chosen to be high (e.g. $c_T=0.95$). Doses also have to be reasonably efficacious to be admissible, such that $p(\pi_E(d) < \phi_E | \textrm{data}) < c_E$, where $\phi_E$ is an efficacy lower limit and $c_E$ is an efficacy threshold, usually chosen to be high (e.g. $c_E=0.9$).

We wish to evaluate two variants of BOIN12 based on different choices of the safety and efficacy thresholds that define the dose-acceptability criteria.  Variant 1 of BOIN12 uses $c_T=0.95$ and $c_E=0.9$, whilst Variant 2 uses $c_T=0.85$ and $c_E=0.8$. We expect these variants of BOIN12 to perform similarly to each other since the designs are identical in all other ways and hence the focus is on small differences in model performance. Other parameter values used to specify the BOIN12 design are given in in Web Table 1.

\subsection{Data generating mechansims}
 We simulate 10,000 PO trials (Set 1) of correlated toxicity (DLT) and efficacy (response) outcomes, as described in Section \ref{section:sim_correlated_pos} with a standard bivariate Gaussian with correlation $\rho=0$. Further simulations are conducted with two different correlation settings ($\rho=-0.5$ and $\rho=0.5$), and we report briefly on these settings at the end of Section \ref{s:case_study_1_results}. Each trial has a maximum sample size of 36, reflecting the size of a typical Phase I/II trial, and patients are dosed in cohort sizes of 3. Five dose levels are considered with true probabilities of DLT equal to $\{0.05,0.10,0.15,0.18,0.45\}$ and true probabilities of response equal to $\{0.40,0.50,0.52,0.53,0.53\}$ (this is the first scenario of the simulation study in Guo and Yuan \cite{guo_droid_2023}). Given the utility values specified in Web Table 1, the true OBD is dose level 2 since it has the highest mean utility and is less than or equal to the maximum tolerated dose (MTD) \cite{lin_boin12_2020}. The outcomes for each of the dose levels is simulated for each patient assuming that both the individual-level toxicity and efficacy outcomes are monotonically non-decreasing. We then simulate a further 10,000 PO trials (Set 2), independent of Set 1, each with maximum sample size of 36.

Variant 1 and 2 of BOIN12 are conducted on the same 10,000 PO trials using Set 1. We also apply the BOIN12 Variant 2 design to the 10,000 independent trials in Set 2. A comparison of BOIN12 Variant 1 using Set 1 with BOIN12 Variant 2 using Set 2 provides an assessment of the difference between methods under a simulation set-up where trials are simulated independently for each design. This comparison resembles the conventional approach where designs are conducted on independently simulated outcomes. Meanwhile a comparison of BOIN12 Variant 1 using Set 1 with BOIN12 Variant 2 using Set 1 provides an assessment of the difference between methods using the same simulated datasets.

\subsection{Monte Carlo Standard Error}
For this demonstration we will focus on one performance measure, the probability of correctly selecting the OBD at the end of the trial. For design $k$ this is defined as $\psi_k = p(\hat{\theta}_k = \theta)$, where $\hat{\theta}_k$ is the estimated OBD from design $k$ and $\theta$ is the true OBD (dose level 2). From a simulation study this is estimated as $\hat{\psi}_k = \frac{1}{n_{sim}} \sum_{i=1}^{n_{sim}} 1(\hat{\theta}_{i,k} = \theta)$. The difference in this performance measure between two competing methods $k=1$ and $k=2$ is $\hat{\delta}_{\psi} = \hat{\psi}_1 - \hat{\psi}_2$ and the (Monte Carlo) variance of the difference is 
\begin{equation}
\textrm{Var}(\hat{\delta}_{\psi}) = \textrm{Var}(\hat{\psi}_1) + \textrm{Var}(\hat{\psi}_2) - 2 \textrm{Cov}(\hat{\psi}_1, \hat{\psi}_2).
\label{eq:MCerror_contrast}\end{equation} 

Let $X_{i} = 1(\hat{\theta}_{i,1}=\theta)$ be the Bernoulli indicator that Design 1 correctly estimates the OBD in simulation $i$ and $Y_{i} = 1(\hat{\theta}_{i,2}=\theta)$ the equivalent from Design 2.
We see that $\textrm{Var}(\hat{\psi}_1) = \frac{1}{n_{sim}} \textrm{Var}(X_i)$ and $\textrm{Var}(\hat{\psi}_2) = \frac{1}{n_{sim}} \textrm{Var}(Y_i)$.

Furthermore $$\textrm{Cov}\left(\hat{\psi}_1, \hat{\psi}_2\right) = \textrm{Cov}\left(\frac{1}{n_{sim}} \sum_{i=1}^{n_{sim}} X_{i},  \frac{1}{n_{sim}} \sum_{j=1}^{n_{sim}} Y_{j}\right) = \frac{1}{n^2_{sim}} \sum_{i=1}^{n_{sim}} \sum_{j=1}^{n_{sim}} \textrm{Cov}(X_i, Y_j) = \frac{1}{n^2_{sim}} \sum_{i=1}^{n_{sim}} \textrm{Cov}(X_i, Y_i)$$
since $\textrm{Cov}(X_i, Y_j) = 0$ for $i \ne j$. If the two designs are evaluated on the same simulated datasets then $\textrm{Cov}(X_i, Y_i) \ne 0$ for $i=1,\ldots,n_{sim}$ and hence there will be an efficiency gain (a reduction in the Monte Carlo standard error) in evaluating the difference in the performance measure between the two designs. 

Plugging in the sample covariance for $\textrm{Cov}(X_i, Y_i)$ and the sample variances for $\textrm{Var}(X_i)$ and $\textrm{Var}(Y_i)$ we can evaluate the Monte Carlo variance in Equation \ref{eq:MCerror_contrast}.

An alternative approach to calculate the Monte Carlo variance in Equation (1) is to calculate it directly using the differences between $X_i$ and $Y_i$ since
$$\hat{\delta}_{\psi} = \frac{1}{n_{sim}} \sum_{i=1}^{n_{sim}} \left( X_i - Y_i \right)$$
and the Monte Carlo variance is then just the sample variance of the differences divided by $n_{sim}$.

\subsection{Simulation results} \label{s:case_study_1_results}
\subsubsection{Correct selection of OBD}
\begin{figure}
	\centering
  \includegraphics[width=\textwidth]{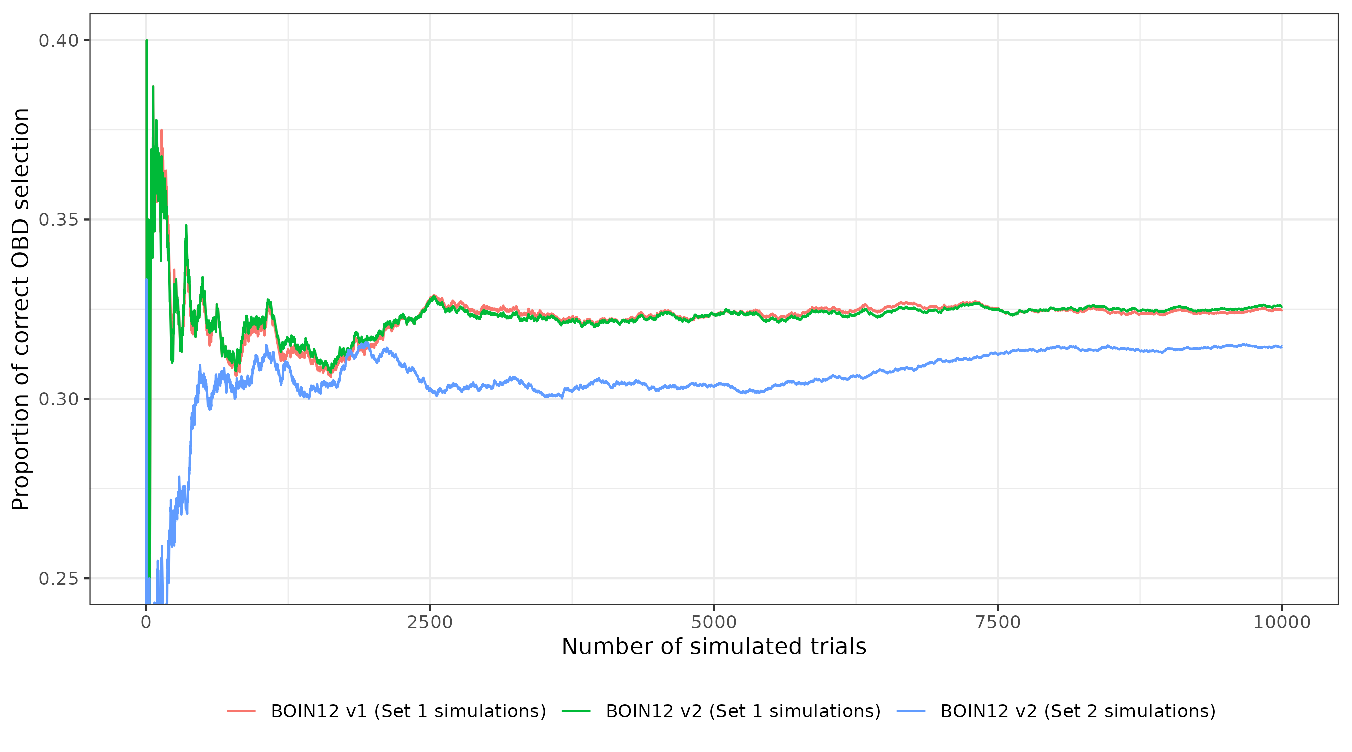}
  \caption{\label{fig:correct_obd_selection} Proportion of trials that correctly select the OBD using BOIN12 Variant 1 and Variant 2 designs and either the same potential outcome trials or independently simulated trials. Results are shown as a function of the number of simulated trials (x-axis).}
	\end{figure}

Figure \ref{fig:correct_obd_selection} shows the estimated proportion of correct OBD selections from the BOIN12 designs based on the accumulating number of simulated trials (x-axis).
Under BOIN12 Variant 1, Set 1 (red line) the proportion of correct OBD selections converges towards 32.5\% after 10,000 simulated trials. BOIN12 Variant 2 applied to the same simulated trials (Set 1) (green line) also converges towards 32.5\% correct OBD selection, whilst BOIN12 Variant 2 applied to independently generated trials (Set 2) (blue line) tracks slightly below 32.5\% after 10,000 trials. Using the same trials (red versus green lines) we see that the proportion of correct selection for BOIN12 Variant 1 is very similar to the BOIN12 Variant 2 performance with the two statistics closely tracking each other since they are applied to the same trials. It is more difficult to compare between the designs when using different trials (red vs. blue lines), and one may erroneously conclude that Variant 2 gives significantly worse performance if we were to only use, say, 5000 simulated trials.

\begin{figure}
	\centering
  \includegraphics[width=\textwidth]{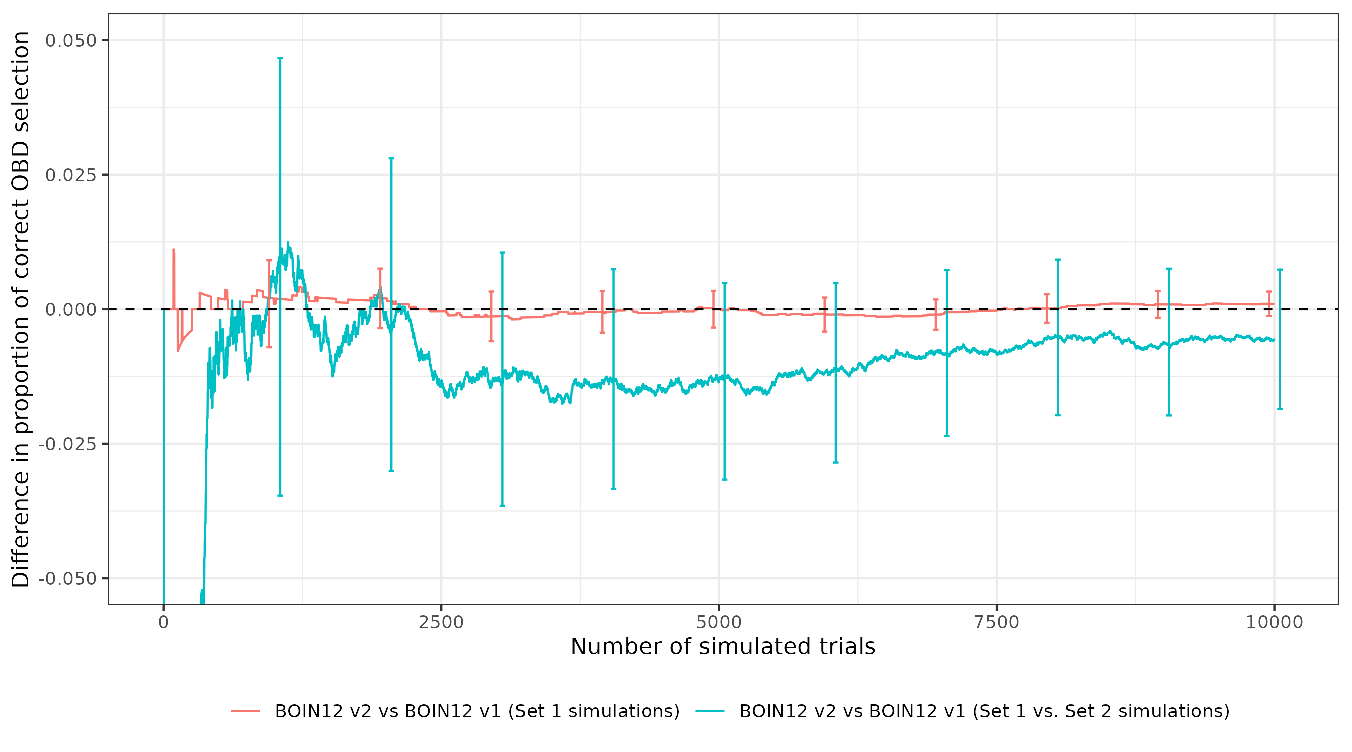}
  \caption{\label{fig:diff_correct_obd_selection} Difference in the proportion of trials that correctly select the OBD using BOIN12 Variant 1 and Variant 2 designs and either the same potential outcome trials or independently simulated trials. Results are shown as a function of the number of simulated trials (x-axis). Error bars indicate +/- 1.96*MCSE calculated every 1000 simulations.}
	\end{figure}

The difference in the proportion of trials that correctly select the OBD between the two variant methods is more clearly seen in Figure \ref{fig:diff_correct_obd_selection}. If independent trials are simulated for the two variants then convergence of the difference is not reached by 10,000 trials, whereas convergence is much more rapid using the same PO trials for the two BOIN12 variants. By 10,000 simulated trials we would conclude that the OBD correct selection percentage is 0.10\% more in Variant 2 than Variant 1 using the same simulated trials, whilst the conclusion using independent trial simulation is that Variant 2 correctly selects the OBD 0.56\% less than Variant 1. Using only 5,000 simulated trials we would have come to a more extreme erroneous conclusion in stating that Variant 2 correctly selects 1.34\% fewer OBDs than Variant 1.

\subsubsection{Monte Carlo Standard Error}

Under independent sampling of simulated datasets the sample correlation of the two estimates is close to zero (-0.01), whilst a substantial positive correlation is evident if the competing methods are applied to the same simulated datasets, especially if the methods often select the same dose. For the BOIN12 case-study, the Pearson correlation of the performance measures under Variant 1 and Variant 2 when applied to the same simulated datasets is 0.97. This gives rise to a Monte Carlo standard error (MCSE) for the difference between the performance measures of 0.00661 using 10,000 independently simulated datasets for each method, and 0.00117 using the same 10,000 simulated datasets for the methods. Put another way, to achieve the same MCSE using independently simulated datasets as obtained using the same simulated datasets we would require $0.00661^2 / 0.00117^2 = 32$ times more simulated trials (e.g. 320,000 simulations compared to 10,000). Given that multiple scenarios are often investigated in a comprehensive simulation study, independently simulated datasets for each scenario would increase computational burden still further. Meanwhile, one set of trials simulated in advance can be used to investigate the performance of a design or multiple designs across multiple design configurations.

\subsubsection{Results for simulations with correlated toxicity and efficacy endpoints}
In the setting where the latent bivariate Gaussian random variables for toxicity and efficacy have a negative correlation ($\rho=-0.5$) the two BOIN-12 variants are much more highly correlated in recommending the same final dose; the performance measures have a correlation of 0.99. This translates into a relative efficiency  of 73 (e.g. 730,000 simulations required using independent simulations compared to 10,000 dependent simulations). Meanwhile under the setting where toxicity and efficacy are positively correlated ($\rho=0.5$) there is less correlation between the performance meaures from the two BOIN-12 variants, with a correlation of 0.94. This translates into a relative efficiency of 18 (e.g. 180,000 simulations required using independent simulations compared to 10,000 dependent simulations).

\section{Case study 2: Comparing a CRM and mTPI-2 design}
\label{s:case_study2}

\subsection{CRM versus mTPI-2}
The second case study applies two common Phase I escalation designs; the Continual Reassessment Method (CRM) \cite{oquigley_continual_1990} and the modified toxicity probability interval (mTPI-2) design \cite{ji_modified_2010}. The aim of these designs is to use a binary toxicity endpoint (DLT) to identify the maximum tolerated dose (MTD). The former design is an example of a model-based design whilst the latter is a model-assisted design. Previous studies have shown that the CRM design generally outperforms the mTPI-2 design but the latter is often preferred due to being operationally simpler to implement.

A scenario with four dose levels is investigated. The target toxicity is 30\% and the true probability of toxicity at each of the dose levels is (0.01, 0.05, 0.15, 0.30). The true MTD is dose level 4 since its true probability of toxicity is equal to the target toxicity (30\%). Implementations of the CRM and mTPI-2 use standard settings as detailed in Web Tables 2 and 3. A maximum of 30 subjects are recruited to the dose escalation study with a stopping rule implemented based on the posterior probability that the lowest dose is too toxic.

\subsection{Data generating mechansims}
As with Case Study 1, we simulate 10,000 PO trials (Set 1) of toxicity (DLT) outcomes. The outcomes for each of the dose levels is simulated for each patient assuming that the individual-level toxicity outcomes are monotonically non-decreasing. We then simulate a further 10,000 PO trials (Set 2), independent of Set 1. We compare CRM versus mTPI-2 when both designs are applied to the same set of PO trials (Set 1) and comparing CRM applied to Set 1 to mTPI-2 applied to Set 2.

\subsection{Simulation results}
\begin{figure}
\centering
\includegraphics[width=\textwidth]{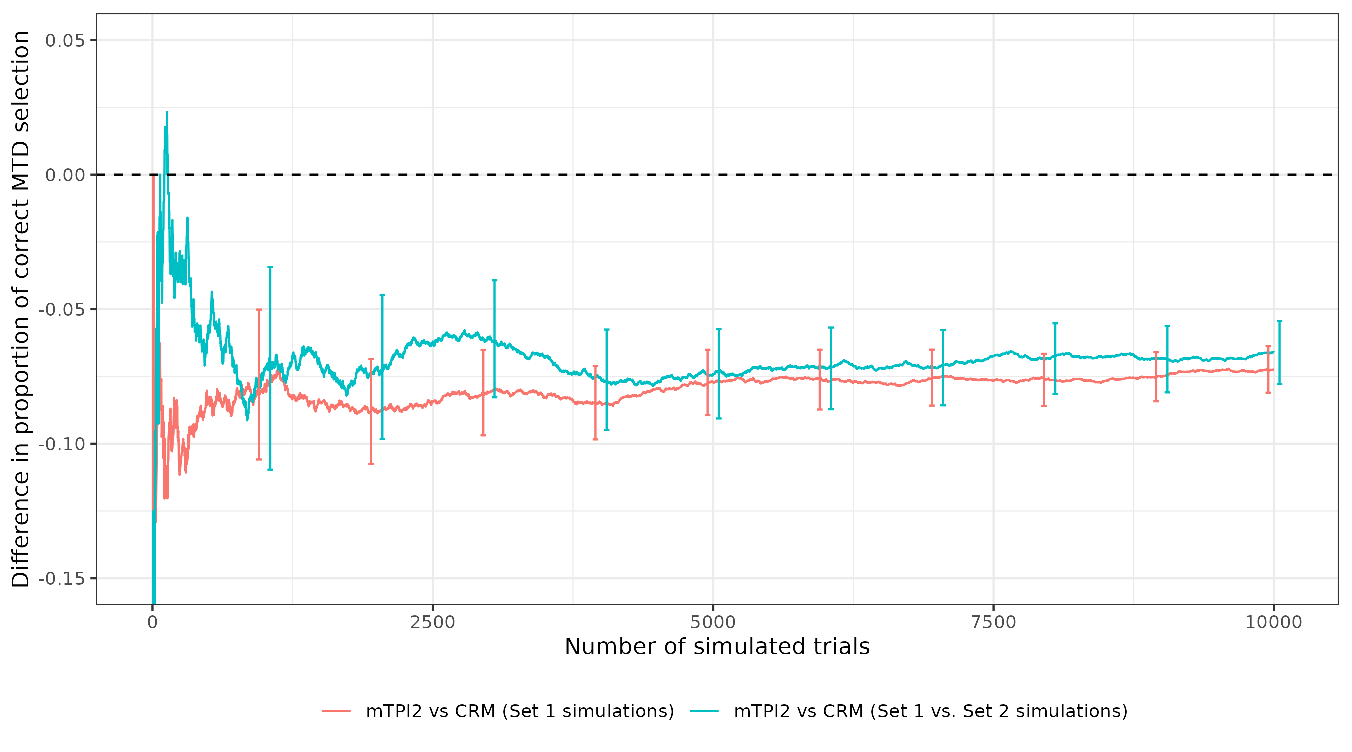}
\caption{\label{fig:diff_correct_mtd_selection} Difference in the proportion of trials that correctly select the MTD using mTPI2 and CRM designs and either the same potential outcome trials or independently simulated trials. Results are shown as a function of the number of simulated trials (x-axis). Error bars indicate +/- 1.96*MCSE calculated every 1000 simulations.}
\end{figure}

After 10,000 simulations using Set 1 the percentage of trials that correctly recommended dose level 4 as the MTD was 81\% using the CRM design compared to 74\% using mTPI-2. The difference in the correct selection percentages between the two designs is shown in Figure \ref{fig:diff_correct_mtd_selection}. Convergence of the difference is achieved after approximately 5,000 simulations when the two designs come from the same set of trials (Set 1; red line), whilst convergence is not quite as rapid when the two designs are applied to different simulated datasets (blue line). The MCSE of the difference, shown by the error bars in Figure \ref{fig:diff_correct_mtd_selection}, is smaller when using common simulated datasets. The MCSE of the difference after 10,000 simulations is 0.004408828 using the same datasets compared to 0.00599472 when using independent datasets. To achieve the same MCSE using independently simulated datasets as obtained using the same simulated datasets we would therefore require $0.00599^2 / 0.00441^2 = 1.85$ times more simulated trials (e.g. 18,500 simulations compared to 10,000).

The efficiency gains seen in Case Study 2 are less than observed in Case Study 1 for a number of reasons. First, the designs in Case Study 2 are based only on one endpoint (a toxicity endpoint) whilst Case Study 1 includes both toxicity and efficacy endpoints. Therefore, there is less sampling (stochastic) variability at any decision stage in Case Study 2. Second, in Case Study 1 a stopping rule was implemented for each of the BOIN12 variants that stopped the trial if 12 or more patients were dosed on the next recommended dose. When the same datasets are used for each of the variants, there is a strong correlation between the designs for when the trial is stopped, leading to a smaller MCSE of the difference in the performance metric. Third, the designs being compared in Case Study 1 are variants of the same escalation design and so are more alike. This translates to small total variance for the difference between designs, and the MCSE is a bigger component of the overall variance, and hence something worth minimising. Whilst in Case Study 2 the MCSE is a small proportion of the total variation as the two designs are quite different.

\section{Implementation}
\label{s:implementation}
All analyses were conducted using the \texttt{escalation} package in R \cite{escalation} (\url{https://cran.r-project.org/web/packages/escalation/index.html}). We have implemented the potential outcome simulation method in the \texttt{simulate\_compare} function in \texttt{escalation}, and a comprehensive vignette describing this function is available at \url{https://brockk.github.io/escalation/articles/A710-SimulationComparison.html}. 
Additional post-simulation functions can produce convergence plots of the simulations simular to those in Figures \ref{fig:correct_obd_selection}-\ref{fig:diff_correct_mtd_selection}, and the latent uniform random variables that manage the occurrence of events for each patient can be exported or imported to the simulation function. Finally, a matrix of potential outcomes can be produced as per Table \ref{tab:po_example} for each simulated trial.

\texttt{escalation} provides considerable flexibility to efficiently compare multiple designs that are implemented within the package, including the 3+3, the continual reassessment method (CRM) \cite{oquigley_continual_1990}, a two-parameter derivative of the CRM \cite{neuenschwander_critical_2008}, the toxicity probability interval (TPI) \cite{ji_dose-finding_2007}, the modified TPI (mTPI-2) \cite{ji_modified_2010}, the Bayesian optimal interval design (BOIN) \cite{liu_bayesian_2015} and BOIN12 \cite{lin_boin12_2020}. A full list of designs can be found in the package vignettes.

Simulation code to replicate results in Case Studies 1 and 2 is provided in the Supplementary Information.

\section{Discussion}
\label{s:discussion}
Simulation studies are an essential component for the evaluation of early phase trial designs but are often time consuming when evaluating many design options under many data generating scenarios. 
Traditionally the design of efficient simulation studies has been overlooked but a properly designed simulation study can lead to substantial time savings whilst estimating operating characteristics to a desired level of accuracy. 
In this paper we introduced the PO simulation paradigm. 
There are many reasons that we recommend this approach becomes the default for simulating Phase I/II data. 
Firstly, PO simulation is relatively simple to perform and implement and hardly more challenging than traditional simulation approaches. 
Secondly, PO simulation vastly reduces the number of simulations required to get to the same desired level of accuracy. 
Thirdly, once simulated, PO datasets can be reused for multiple evaluations and can be shared between the trial team allowing a more comprehensive evaluation of individual simulated trials where operating characteristics are suboptimal. 
An individual PO simulated trial can be further interrogated to understand where decisions between two competing designs deviate. 
Furthermore, through sharing of simulated PO datasets or the algorithm and random number state used to generate them, operating characteristics can be independently verified by other researchers and results exactly reproduced. Indeed, the same PO datasets can further be applied by other researchers to new designs. 
Finally, to aid use we have implemented the approach in the R package \texttt{escalation} \cite{escalation} 
via the \texttt{simulate\_compare} function.
The core aspect of this function is that it uses the same notional patients within each simulated trial across the designs being compared. 
Another important feature is that it permits the return, storage and importing of the latent variables that underpin PO simulated datasets. 
\texttt{escalation} supports a common inferface and provides considerable flexibility for users to daisy-chain together design options. 
We recommend, and indeed implement, the use of "convergence" plots, such as those produced in Figures \ref{fig:correct_obd_selection} and \ref{fig:diff_correct_obd_selection}, to visually understand MC error in simulation studies.
Further information on simulating with \texttt{escalation} are publicly available at https://brockk.github.io/escalation/.

Despite its numerous advantages, the PO simulation approach has some limitations. Firstly, existing software must be re-engineered to change the way that the data are simulated. Currently, if a user has simulated PO datasets, they cannot easily be passed to existing software for evaluation of the design. We have anticipated this limitation and have modified the \texttt{escalation} package to both simulate or accept PO datasets, and also to return them to the user if requested. We hope other software engineers will follow suit. Secondly, the PO simulation approach only reduces MC error for contrasts of competing designs. The new framework will not help in reducing the number of simulation iterations required to achieve the desired level of accuracy for the absolute value of a performance measure of a given design. Nevertheless, the PO simulation approach can be utilised to select a superior design out of a set of competing designs. Once this exercise is done, the simulation study may be enlarged to get accurate estimates of absolute measures for the final design. 

An alternative approach often proposed to making a simulation study more efficient and hence reducing MC error is to “set the random seed” at the beginning of a simulation to be the same when evaluating competing designs. However, this approach breaks down as soon as the two designs deviate in their recommendation and pseudo-random number chains can become misaligned. An example of this is when multiple trials are being simulated. The first trial may start at the same location of the pseudo-random number chain for Design 1 and Design 2, due to the same “seed” being given. However, if Design 1 stopped early at patient $m<N$ whilst Design 2 continued to patient $N$ then for the next simulated trial, the random number chains will be misaligned and the generated outcome for patient 1 of the trial will use pseudo random number $m+1$ under Design 1 and pseudo random-number $N+1$ under Design 2. This introduces unwanted stochastic variation when comparing between designs.

Morris et al. give a comprehensive discussion of the calculations needed to choose a suitable number of iterations for a simulation study \cite{morris_using_2019}. If the interest is in the contrast of a performance measure between two designs then an estimate of the between-design correlation is required when applied to PO simulated trials. This may be difficult to know a-priori and hence a small initial PO simulation might be undertaken to empirically obtain an estimate of the correlation. Then formulae such as those in Equation \ref{eq:MCerror_contrast} can be rearranged to get an estimate of $n_{sim}$ for a desired MC error. 

We have not considered continuous efficacy endpoints in this paper though an extension to this setting is rather trivial. As in the binary endpoint setting, a patient-specific uniform random variable can be generated and used to obtain the continuous potential outcomes for each dose level using an appropriate inverse cumulative distribution function, with each dose level having its own mean and variance. 

In summary, we have introduced a new PO framework for reducing computational and analyst time when conducting simulation studies for Phase I/II designs. We recommend it is considered widely for future implementation of simulation studies and is the default approach when comparing competing methods. 

\backmatter 

\section*{Acknowledgements}
MS, DS, and DJ are full-time employees of AstraZeneca and declare AstraZeneca stock ownership. KB is a contracted employee of AstraZeneca and declares GlaxoSmithKline stock ownership.

\bibliographystyle{biom}
\bibliography{Potential_Outcome_Simulation}

\section*{Supporting Information} Web Tables 1-3, referenced in Sections \ref{s:case_study1} and \ref{s:case_study2}, is available under the Paper Information link at the \textit{Biometrics} website \texttt{http:www.tibs.org/biometrics}. Simulation code to replicate results in Case Studies 1 and 2 is provided in the Supplementary Information.

\label{lastpage}

\end{document}


\label{firstpage}

\begin{keywords}
Dose-finding, adaptive designs, simulation, MCMC error, potential outcomes
\end{keywords}

\maketitle

\begin{table}
\caption{\label{tab:BOIN12_parameter_values} Parameter values used to specify the BOIN12 designs compared in Case Study 1.}
\begin{center}
\begin{tabular}{cc}
BOIN12 Parameter & Value \\
\hline
Cohort size & 3 \\
Maximum sample size & 36 \\
Maximum number treated per dose & 12 \\
$N^*$ (sample size cut-off for active exploration of new doses) & 6 \\
$\phi_T$ (upper limit for DLT) & 0.35 \\
$\phi_E$ (lower limit for response) & 0.25 \\
$u_2$ (utility ascribed to no DLT, no response outcome) & 40 \\
$u_3$ (utility ascribed to DLT, response outcome) & 60 
\end{tabular}
\end{center}
\end{table}

\begin{table}
\caption{\label{tab:CRM_parameter_values} Parameter values used to specify the CRM design in Case Study 2.}
\begin{center}
\begin{tabular}{cc}
CRM Parameter & Value \\
\hline
Cohort size & 3  \\
Maximum sample size & 30 \\
Target toxicity & 0.30 \\
Skeleton probabilities & (0.05, 0.15, 0.30, 0.45)  \\
Model & Logistic \\
Intercept & 3 \\
Prior Standard Deviation & $\sqrt{1.34}$ \\
Stopping Rule & $p(\texttt{Tox rate at dose 1} > 0.3) > 0.8$
\end{tabular}
\end{center}
\end{table}

\begin{table}
\caption{\label{tab:mTPI-2_parameter_values} Parameter values used to specify the mTPI-2 design in Case Study 2.}
\begin{center}
\begin{tabular}{cc}
mTPI-2 Parameter & Value \\
\hline
Cohort size & 3  \\
Maximum sample size & 30 \\
Target toxicity & 0.30 \\
Equivalence interval & (0.30 - 0.05, 0.30 + 0.05) \\
Priors & $Beta(0.5, 0.5)$ \\
Inadmissible dose & $p(\texttt{Tox rate at dose} > 0.3) > 0.95$
\end{tabular}
\end{center}
\end{table}

\label{lastpage}